\documentclass[
prb,twocolumn,amsmath,amssymb,10pt,aps
,longbibliography,superscriptaddress,citeautoscript,bibnotes
,floatfix
]{revtex4-2}

\usepackage{comment}
\usepackage{dsfont}
\usepackage{color}
\usepackage{tikz}
\usepackage{xcolor}
\usepackage{colortbl}
\usepackage{amsthm}
\usepackage{braket}
\usepackage[breaklinks]{hyperref}
\usepackage{longtable}
\usepackage{xtab}
\usepackage{cleveref}
\usepackage{float}
\usepackage{mathtools}

\newcommand{\ra}{\rightarrow}

\newcommand{\ZZ}{{\mathbb Z}}

\newcommand{\cB}{\mathcal B}
\newcommand{\cC}{\mathcal C}
\newcommand{\cD}{\mathcal D}
\newcommand{\cL}{\mathcal L}

\newcommand{\cT}{{\mathcal T}}

\newcommand{\hS}{\hat{S}}
\newcommand{\hT}{\hat{T}}

\newcommand{\SL}{\mathrm{SL}_2(\mathbb{Z})}
\newcommand{\Gth}{\Gamma_\theta}

\newcommand{\SLn}{\mathrm{SL}_2(\mathbb{Z}_N)}

\newcommand{\cS}{\mathcal{S}}

\begin{document}

\title{Gapped boundaries of fermionic topological orders and  higher central charges}
\author{Minyoung You}
\email{miyou849@gmail.com}
\affiliation{POSTECH Basic Science Research Institute, Pohang, Gyeongbuk, 37673, Korea}

\begin{abstract}
We develop a test for the vanishing of higher central charges of a fermionic topological order, which is a necessary condition for the existence of a gapped boundary, purely in terms of the modular data of the super-modular tensor category. More precisely, we test whether a given super-MTC has $c = 0$ mod $\frac{1}{2}$, and, if so, whether the modular extension with $c =0$ mod $8$ has vanishing higher central charges. The test itself  does not require an explicit computation of the modular extensions and is easily carried out. We apply this test to known examples of super-modular tensor categories. Since our test allows us to obtain information about the chiral central charge of a super-modular tensor category in terms of its modular data without direct knowledge of its modular extensions, this can also be thought of as the first step towards a fermionic analogue of the Gauss-Milgram formula.
\end{abstract}

\maketitle

\section{Introduction}
\label{sec:intro}

Gapped boundaries of $2+1$d topological orders have been  studied extensively. The low-energy limit of topological orders are described by $2+1$d topological quantum field theories (TQFT), which in turn have a mathematical description in terms of modular tensor categories (MTC) \cite{Bakalov2000LecturesOT, Kitaev_2006, wang2010topological, RevModPhys.80.1083}. Thus, a systematic study of gapped boundaries of $2+1$d topological orders have been carried out in terms of the data of MTCs, or, equivalently, in terms of the anyonic excitations of topological orders and their braiding and fusion properties 
\cite{Kapustin_2011, davydov2011witt, Kitaev_2012, Fuchs_2013, Kong_2014, Lan_2015}. In particular, it was shown that a topological order admits a gapped boundary if and only if it has a set of anyons with topological spin $\theta_a =1$ with trivial mutual braiding with each other which can \emph{condense}, or, mathematically, form a Lagrangian algebra object of the MTC \cite{Bais_2009, Kong_2014, Burnell_2018, Lou_2021}. 

In general, it is difficult to determine if a given MTC $\cC$ has a Lagrangian algebra. 
 There are, however, alternative quantities which can easily computed from  the $S$- and $T$-matrices of the MTC (collectively called the \emph{modular data}), which provide obstructions to the existence of a gapped boundary. 
 
The best known of these is the chiral central charge. It can be computed through the \emph{Gauss-Milgram formula}
\begin{equation}
    e^{2 \pi  i c/8} = \frac{1}{D} \sum_a d_a^2 \theta_a
    \label{eq:Gauss_Milgram}
\end{equation}
where $d_a = \frac{S_{1a}}{S_{11}}$ are the quantum dimensions and $\theta_a = T_{aa}$ are the topological spins of the anyons $a$. Eq. \ref{eq:Gauss_Milgram} shows that the bulk anyon data determines the chiral central charge mod $8$, and whenever $c \neq 0$ mod $8$, the bulk data is inconsistent with the existence of a gapped boundary.

Even when $c = 0$, however, there may not exist any gapped boundary  \cite{Levin_2013}. 
We can also define additional quantities called \emph{higher central charges}:
\begin{equation}
    \xi_n = \frac{\sum_a d_a^2 \theta_a^n}{|\sum_a d_a^2 \theta_a^n|}
    \label{eq:hcc}
\end{equation}
Existence of a gapped boundary implies $\xi_n = 1$ for all $n$ such that ${\rm gcd}(n, N_{FS}) = 1$ \cite{Ng_2019, Ng_2022, Kaidi_2022}.
Here, $N_{FS}$ is the Frobenius-Schur exponent, defined the order of the $T$-matrix. While the quantity Eq. \ref{eq:hcc} is defined for any $n$, in the rest of the paper by ``higher central charge'' we will refer to $\xi_n$ for $n$ coprime to $N_{FS}$, since those are the quantities relevant to the existence of a gapped boundary \footnote{For abelian MTCs,  more general values  of $n$ are known to give obstructions to the existence of a gapped boundary, and in fact taken together they form a sufficient condition for the existence of a gapped boundary \cite{Kaidi_2022}. In this paper, we focus on the general non-abelian case, and only consider $\xi_n$ for $n$ coprime to $N_{FS}$. Indeed, the abelian case will not be of much interest for us, since abelian fermionic topological orders are never ``intrinsically fermionic'': they can always be obtained by stacking a bosonic topological order with the trivial fermionic topological order \cite{abelian_split, cho2023classification}}.   Moreover, we will say that a higher central charge $\xi_n$ \emph{vanishes} if it satisfies $\xi_n = 1$, since it is defined multiplicatively (unlike $c$ which is defined additively).

It is natural to ask whether there are analogous quantities for fermionic topological orders -- topologically-ordered phases of systems with fermions,  whose bulk excitations are described by super-modular tensor categories (super-MTC) \cite{Lan_2016theory, Bruillard_2017}, and which give rise to spin-TQFTs at low energies \cite{Gaiotto_2016, Bhardwaj_2017}. In fact, the Gauss-Milgram formula Eq. \ref{eq:Gauss_Milgram} has  no straightforward fermionic analogue, because the sum  is identically zero for a super-MTC \cite{Lan_2016theory}. This is related to the fact that the $S$-matrix of a super-MTC is degenerate.  
Hence, traditionally, in order to define the chiral central charge $c$ for a super-MTC $\cB$, we first take the modular extension, which is an MTC $\breve{\cB}$ which contains $\cB$ as a subcategory \cite{Lan_2016theory, Lan_2016me, Bruillard_2017, Johnson_Freyd_2023}. There are $16$ possible modular extensions, with different chiral central charges; however, the chiral central charges are the equivalent mod $\frac{1}{2}$, so the chiral central charge of a super-MTC is well-defined mod $\frac{1}{2}$ \cite{Cano_2014, Lan_2016theory, cho2023classification}. 

When $\cB$ has  $c =0$ mod $\frac{1}{2}$, $\cB$ will admit a modular extension $\breve{\cB}$ with $c = 0$ mod $8.$ However, even then, $\breve{\cB}$  may not admit a gapped boundary.   For the purpose of the main text, we will assume the following:
\begin{itemize}
    \item  \emph{ A super-MTC $\cB$ admits a  gapped boundary if and only if has a  modular extension  $\breve{\cB}$ which admits a gapped boundary. }
\end{itemize}
While the theory of gapped boundaries for fermionic topological orders in terms of Lagrangian algebras is yet to be fully developed (though see Refs. \cite{Wan_2017,  Kobayashi_2022, chen2022boundaries, Lou_2021}), this assumption is consistent with all known examples of anyon condensation from $\cB$ to the trivial super-MTC \cite{Wan_2017}.  Moreover,  in Appendix \ref{app:gapped_bc} we will show that a gapped boundary for modular extension on the torus can be used to construct a gapped boundary for the corresponding $2+1$d spin-TQFT on the torus and vice versa.

Since $c = 0$ mod $\frac{1}{2}$ is a necessary condition for $\cB$ to admit a $c=0$ mod $8$ modular extension, this is equivalent to saying that the fermionic topological order $(\cB, c)$ admits a gapped boundary if and only if $c =0$ and the and the unique modular extension with $c =0$, which we denote as $(\breve{\cB}, 0)$, admits a gapped boundary. 
Then, the vanishing of higher central charges Eq. \eqref{eq:hcc} for $(\breve{\cB},0)$ becomes a necessary condition for the fermionic topological order $(\cB,0)$ to admit a gapped boundary. 

The above discussion may suggest that, in order to test whether a given super-MTC $\cB$ admits a gapped boundary, we first need to compute the modular extension $(\breve{\cB}, 0)$ and then compute its higher central charges. The main result of this paper, however, is that this is not the case: we can in fact compute whether or not the higher central charges of the modular extension vanish purely in terms of the modular data of the super-MTC $\cB$, without having to explicitly compute the modular extension first.

It is important to note that, while the higher central charges $\xi_n$ are well-defined up to factors of $e^{2 \pi i \frac{1}{16}}$ for a given super-MTC $\cB$ (as we will show  in Appendix \ref{app:hcc_defined}), the vanishing of $\xi_n$ modulo  this factor does not guarantee that  $(\breve{\cB}, 0)$ will have vanishing $\xi_n$. This is because, if, for example $\xi_n = -1$ for some $n$ coprime to $N_{FS}$, it is $1$ modulo factors $e^{2 \pi i \frac{1}{16}}$, but is nevertheless not actually $1$.  Our test determines, for the case $c = 0$, all the higher central charges $\xi_n$ simultaneously, so relative phases between the higher central charges are taken into account.

In Sec. \ref{sec:congruence}, we review the notion of congruence representations of the modular group $\SL$, which will be crucial in testing whether a given super-MTC has $c =0$ mod $\frac{1}{2}$. In Sec. \ref{sec:super_MTC}, we review the structure of modular extensions of super-MTCs, and show that the order of the $T$-matrix of the modular extension can be computed from the modular data of super-MTCs.  In Sec. \ref{sec:test}, we state our test and prove that it indeed tests $c = 0$ mod $\frac{1}{2}$ and the vanishing of higher central charges of the $c =0$ modular extension. In Sec. \ref{sec:examples}, we summarize which known super-MTCs pass this test and hence potentially admit gapped boundaries.

\section{Congruence representations of the modular group}
\label{sec:congruence}

It is well-known that the modular data $(S, T)$ of an MTC form a projective  unitary representation 
of the modular group $\SL$, generated by 
\begin{equation}
    s = \begin{pmatrix}
        0 & - 1 \\  1 & 0
    \end{pmatrix}, \ t = \begin{pmatrix}
        1 & 1 \\ 0 & 1 
    \end{pmatrix},
\end{equation}
where dimension of the representation $r$ equals the rank of the MTC. In fact, the representation also has to be \emph{congruence}, i.e. the kernel of $\tilde{\rho}$ contains $\Gamma(N)$, the principal congruence subgroup of level $N$ of $\SL$, defined as
\begin{align}
    \Gamma(N) &\nonumber \\ 
    =&  \Set{\begin{pmatrix}
        a & b \\ c & d 
    \end{pmatrix} \in \SL | \begin{pmatrix}
        a & b \\ c & d 
    \end{pmatrix} \equiv \begin{pmatrix}
        1 & 0 \\ 0 & 1 
    \end{pmatrix} \text{mod } N}.
\end{align}
The smallest $N$ for which $\ker \rho \geq \Gamma(N)$ is called the level of $\Tilde{\rho}$ \cite{Ng_2010}.  
The lift of these projective representations to linear representations $\rho$ as
\begin{align}
    \rho(s) &:= S \nonumber \\ 
    \rho(t) &:= e^{-2 \pi i c/24} T
    \label{eq:linear_lift}
\end{align}
is  congruence, though of a different level in general \cite{Dong_2015}. Equivalently, congruence representations can be defined as those representations  of $\SL$ which can also be thought of as representations of $\SL/\Gamma(N) \simeq \SLn$.   

Concretely, we can characterize congruence representations in terms of the  relations they need to satisfy in addition to the usual relations of $\SL$ such as $(\rho(s) \rho(t))^3 = \rho(s)^2$ and $\rho(s)^4 = \mathds{1}$. We first define
\begin{align}
    H(n) :=& \rho\begin{pmatrix}
        n & 0 \\ 0& \bar{n} 
    \end{pmatrix}   \nonumber \\ 
    =& \rho(s)^2 \rho(t)^{n^2-n} \rho(s) \rho(t)^{-(\bar{n}-1)} \rho(s) (\rho(t)^2 \rho(s))^{n-1}
    \label{eq:Hn}
\end{align}
where $n \in \ZZ_{N}^\times$ and $\bar{n}$ satisfies $n \bar{n} = 1 \mod N$ \footnote{The expression in terms of $\rho(s)$ and $\rho(t)$ is slightly different from the standard expression  in e.g. Refs. \cite{Dong_2015, Harvey_2018}. We use this equivalent expression, first  obtained in Ref. \cite{cho2023classification}, since this generalizes to $\Gth$.}. Note that $H(n)$ is the same quantity that is used in the Galois conjugation of the $S$-matrix, and is always a signed permutation matrix \cite{Dong_2015, Harvey_2018}. A congruence representation then has to satisfy \cite{Eholzer_1995}:
\begin{align}
    \rho(t)^N &= \mathds{1} \nonumber \\ 
        \rho(s)^2 &= H(-1) \nonumber \\ 
        H(n_1) H(n_2) &= H(n_1 n_2) \nonumber \\ 
        H(n) \rho(t) &= \rho(t)^{n^2} H(n) \nonumber\\
        \rho(s) H(n) &= H(\bar{n}) \rho(s)
        \label{eq:congruence}
\end{align}
for all $n_1, n_2, n, \bar{n} \in \ZZ_N^\times.$
A projective congruence representation will satisfy these relations up to a phase. Note that the relations depend on the level $N$ instead of being universal like $(\rho(s) \rho(t))^3 = \rho(s)^2.$

It was shown in Ref. \cite{Bonderson_2018}
 that the projective representation $\Tilde{\hat{\rho}}$ of $\Gth$ form by the modular data of a super-MTC is also congruence. Moreover, Ref. \cite{cho2023classification} showed that  the linear lift 
 \begin{align}
     \hat{\rho}(s) &:= \Tilde{\hat{\rho}}(s)  \nonumber \\      
     \hat{\rho}(t^2) &:= e^{-2 \pi i c/12}\Tilde{\hat{\rho}}(t^2) 
\label{eq:linear_rep_Gth}
 \end{align}
 (where $c$ is the central charge, defined mod $\frac{1}{2}$, of the super-MTC) is also a congruence representation of $\Gth$. This means that, if we define 
\begin{align}
    \hat{H}(n) :=& \hat{\rho}\begin{pmatrix}
        n & 0 \\ 0& \bar{n} 
    \end{pmatrix}   \nonumber \\ 
    =& \hat{\rho}(s)^2 \hat{\rho}(t)^{n^2-n} \hat{\rho}(s) \hat{\rho}(t)^{-(\bar{n}-1)} \hat{\rho}(s) (\hat{\rho}(t)^2 \hat{\rho}(s))^{n-1}
    \label{eq:hatHn}
\end{align} 
 (note that, because the level $N$ is always even for a congruence representation coming from a super-MTC, any $n \in \ZZ_N^\times$ is odd and the expression Eq. \eqref{eq:hatHn} is well-defined), the representation $\hat{\rho}$ has to satisfy  
\begin{align}
    \hat{\rho}(t)^N &= \mathds{1} \nonumber \\ 
        \hat{\rho}(s)^2 &= \hat{H}(-1) \nonumber \\ 
        \hat{H}(n_1) \hat{H}(n_2) &= \hat{H}(n_1 n_2) \nonumber \\ 
        \hat{\rho}(s) \hat{H}(n) &= \hat{H}(\bar{n}) \hat{\rho}(s).
    \label{eq:congruence_hat}
\end{align}
Compared to Eq. \eqref{eq:congruence}, the condition inovolving $\rho(t)$ does not exist since it does not belong to a representation of $\Gth$. We also note that $N$ is not necessarily the order of $\hat{\rho}(t).$

\section{Super-MTC and modular extensions}
\label{sec:super_MTC}

\subsection{Review of the structure of modular extensions}
\label{sec:super_MTC_MEX}

The modular data of super-MTC always admit the following tensor decomposition \cite{Bruillard_2017}:
\begin{equation}
    S = \frac{1}{2}
    \begin{pmatrix}
        1 & 1 \\
        1 & 1 
    \end{pmatrix} 
    \otimes \hat{S}, 
    \quad T = 
    \begin{pmatrix}
        1 & 0 \\
        0 & -1
    \end{pmatrix}
    \otimes \hat{T}.
\end{equation}
Note that $\hT$ is not canonical, but $\hT^2$ is well-defined. While $S$ is degenerate, $\hS$ is unitary, and $\hS$ and $\hT^2$ together define a representation of $\Gth$, which is a subgroup of $\SL$ generated by $s$ and $t^2.$ 

Given a super-MTC $\cB$, a modular extension $\breve{\cB}$ is an MTC (i.e. it has a nondegenerate $S$-matrix) which contains $\cB$ as a subcategory. Modular extensions with the smallest global dimensions $D_{\breve{\cB}}^2$ are called minimal modular extensions, and in fact they have $D_{\breve{\cB}}^2 = 2 D_{\cB}^2$. In this paper, ``modular extension'' always refers to minimal modular extensions. 

It is known that a modular extension always exists \cite{Johnson_Freyd_2023}, and there are always $16$ different modular extensions  \cite{Lan_2016me}. $\breve{\cB}$ is naturally a \emph{spin-modular tensor category} (spin-MTC): i.e. an MTC with a distinguished fermion $f$, which can be condensed to obtain the super-MTC $\cB$ \cite{Bruillard_2017}. The presence of $f$ imposes the following structure on the modular data of modular extensions:
\begin{itemize}
    \item Each anyon $a$ has mutual braiding phase $\epsilon_a = \pm 1$ with $f$, and we can divide the set of anyon types into $\Pi_{\rm NS}$ (those anyons with $\epsilon_a = +1$) and $\Pi_{\rm R}$ (those anyons with $\epsilon_a = -1$).

    \item Each  $a \in \Pi_{\rm NS}$ satisfies $a \times f \neq a$. We can then (non-canonically) divide $\Pi_{\rm NS}$ into two sets $\Pi_{\rm NS}^1$ and $f \Pi_{\rm NS}$ of equal size. 

    \item For $a \in \Pi_{\rm NS}$, it may satisfy either $a \times f \neq a$ or $a \times f = a$. In the former case,

    \item In the basis given by $\Pi_{\rm NS}^1 \cup f \Pi_{\rm NS}^1   \cup \Pi_{\rm R}^1 \cup \Pi_{\rm R}^f \cup \Pi_{\rm R}^\sigma$, the modular data take the form 
    \begin{align}
    S &= 
    \begin{pmatrix}
        \frac{1}{2}\hat{S} & \frac{1}{2}\hat{S} & A & A & X \\
        \frac{1}{2}\hat{S} & \frac{1}{2}\hat{S} & -A & -A & -X \\
        A^T & -A^T & B & -B & 0 \\
        A^T & -A^T & -B & B & 0 \\
        X^T & -X^T & 0 & 0 & 0
    \end{pmatrix}, \nonumber \\
    T &=
    \begin{pmatrix}
        \hat{T} & 0 & 0 & 0 & 0 \\
        0 & -\hat{T} & 0 & 0 & 0 \\
        0 & 0 & \hat{T}_v & 0 & 0 \\
        0 & 0 & 0 & \hat{T}_v & 0 \\
        0 & 0 & 0 & 0 & T_\sigma
    \end{pmatrix}.
    \label{eq:MEX_ST}
    \end{align}

\item Let consider basis vectors labeled  by anyons as $|a \rangle$, which correspond to a  basis of states of the Hilbert space on the torus of the $2+1$d TQFT defined by the MTC $\breve{\cB}$. We can form a different basis given by 
\begin{align}
    \Pi_{\text{NS-NS}} &= \left\{ \frac{1}{\sqrt{2}} \left(|a\rangle + |a \times f\rangle \right) : a \in \Pi_{\rm NS}^1 \right\} \nonumber \\
    \Pi_{\text{NS-R}} &= \left\{ \frac{1}{\sqrt{2}} \left(|a\rangle - |a \times f\rangle \right) : a \in \Pi_{\rm NS}^1 \right\} \nonumber \\
    \Pi_{\text{R-NS}} &= \left\{ \frac{1}{\sqrt{2}} \left(|a\rangle + |a \times f\rangle \right) : a \in \Pi_{\rm R}^1 \right\} \nonumber \\ 
    & \ \ \cup \left\{ |a \rangle : a \in \Pi_{\rm R}^\sigma \right\} \nonumber \\
    \Pi_{\text{R-R}} &= \left\{ \frac{1}{\sqrt{2}} \left(|a\rangle  - |a \times f\rangle \right) : a \in \Pi_{\rm R}^1 \right\}.
    \label{eq:basis}
\end{align}
These states correspond to the states in a definite spin structure sector in the spin-TQFT obtained from a fermion condensation from $\breve{\cB}.$ In this basis, $S$ and $T$ take the form
\begin{align}
    S = S_I \oplus S_{\text{R-R}} \nonumber \\ 
    T = T_I \oplus T_{\text{R-R}}
\label{eq:RRreduce}
\end{align}
    where
\begin{align}
           S_I &= 
    \begin{pmatrix}
        \hS & 0 & 0 \\
        0 & 0& \hS' \\
        0 & \hS'^T & 0    
    \end{pmatrix}, \nonumber \\
    T_I &= \begin{pmatrix} 0 & \hT  & 0  \\ \hT & 0&  0  \\ 0 & 0 & T_{\text{R-NS}}  \end{pmatrix}.
    \label{eq:spinsector}
\end{align}
Here, 
\begin{align}
    S' &= \begin{pmatrix}
        2A & X 
    \end{pmatrix} \nonumber \\ 
    T_{\text{R-NS}} &=  \begin{pmatrix}
        \hT_v & 0 \\ 0& T_\sigma
    \end{pmatrix} 
    \label{eq:spinsector_RNS}
\end{align}
    are  $r \times r$-matrices, and
    \begin{align}
         S_{\text{R-R}} = 2B \nonumber \\ 
    T_{\text{R-R}} = \hT_v
    \label{eq:spinsector_RR}
    \end{align}
are $r_v \times r_v$-matrices, where $r_v$ is the number of anyons in $\Pi_{\rm R}^1.$
\end{itemize}
We refer to Refs. \cite{Bruillard_2017, Bonderson_2018, Delmastro_2021} for proof of these statements. 

\subsection{Order of \texorpdfstring{$T$-matrix}{} from  super-MTC data} 
\label{sec:order}

Given an MTC $\cC$, the projective representation of $\SL$ defined by $S$ and $T$  can always be lift to a linear representation $\rho$ by Eq. \eqref{eq:linear_lift}. As with its projective counterpart (Eqs. \eqref{eq:RRreduce} and \eqref{eq:spinsector}), this is a reducible representation 
\begin{align}
    \rho = \rho_I \oplus \rho_{\text{R-R}}
    \label{eq:rho_rhoI}
\end{align}
where $\rho_I(s) = S_I$ and $\rho(t) = e^{-2 \pi i c/24} T_I.$ Physically, $\rho_I$ is the representation of the states of the first three sectors (NS-NS, NS-R, R-NS)  of the spin-TQFT under modular transformations \cite{Delmastro_2021} (or, equivalently, it is the representation of the NS-NS, NS-R, and R-NS sector characters of the  fermionic rational conformal field theory  living on the boundary \cite{Duan_2023}).

On the other hand, given the $\Gth$-representation $\hat{\rho}$, we can compute the action of modular transformations on the NS-R and R-NS sectors, using the fact that the NS-R sector states are obtained by a $t$-transformation and R-NS sector states are obtained by  an $st$-transformation on the NS-NS sector states. We can then compute the resulting $\SL$-representation  
\begin{align}
     \rho^{\rm ind}(s) &= 
    \begin{pmatrix}
        \hat{\rho}(s) & 0 & 0 \\
        0 & 0 & \hat{\rho}(s)^2 \\
        0 & \mathds{1} & 0
    \end{pmatrix}, \nonumber \\ 
    \rho^{\rm ind}(t) &=
    \begin{pmatrix}
        0 & \hat{\rho}(t)^2 & 0 \\
        \mathds{1} & 0 & 0 \\
        0 & 0 & (\hat{\rho}(s)\hat{\rho}(t)^2)^{-1}
    \end{pmatrix}.
\end{align}
This is nothing but the representation of $\SL$ \emph{induced} from the representation $\hat{\rho}$ of the subgroup $\Gth$ \cite{cho2023classification} \footnote{While $\SL$ and $\Gth$ are infinite groups, since $\hat{\rho}$ is congruence we can think of it as a representation of the finite group $\Gth/\Gamma(N)$ for some $N$, and compute the representation of the finite group $\SL/\Gamma(N)$ induced from its index $3$ subgroup $\Gth/\Gamma(N)$. Thus the induction is straightforward. }. Hence we expect $\rho^{\rm ind}$ to be a representation equivalent to $\rho_I$. We prove this fact in Appendix \ref{app:induced}. 

Now, we note that by Eqs. \eqref{eq:spinsector}, \eqref{eq:spinsector_RNS}, and \eqref{eq:spinsector_RR}, the entries of $T_{\text{R-R}}$ simply duplicate some of the entries in $T_{\text{R-NS}}$, and hence ${\rm ord} \rho(t)= {\rm ord}\rho_I(t)$ (note that $(\rho_I(t))^2$ is diagonal). On the other hand, since $\rho_I$ is equivalent to $\rho^{\rm ind}$, given $\hat{\rho}$ we can simply compute the eigenvalues of $\rho^{\rm ind}$  to get a diagonal matrix which and find the its order. This will be equal to the order of $\rho(t).$

When $c =0$ mod $\frac{1}{2}$, we can make the choice $c =0$ everywhere so that $\hat{\rho}(t^2) = \hT^2$  and $\rho(t) = T$.  we have  
\begin{align}
    \rho^{\rm ind}  = T^{\rm ind} =
    \begin{pmatrix}
        0 & \hat{T}^2 & 0 \\
        \mathds{1} & 0 & 0 \\
        0 & 0 & (\hat{S}\hat{T}^2)^{-1}
    \end{pmatrix}
    \label{eq:ind_superMD}
\end{align}
and the order of $T$ is equal to the order of the eigenvalues of $T^{\rm ind}.$

\section{Test for vanishing of higher central charges}
\label{sec:test}

We first state the procedure, then prove that this works. 

\subsection*{The test}
We are given the modular data $\hS, \hT^2$ of a super-MTC $\cB$. We wish to test 1) if $\cB$ has vanishing $c = 0$ mod $\frac{1}{2}$ and 2) if so, whether the modular extension $(\breve{\cB}, 0)$ has vanishing higher central charges. We do this in two steps. 

\begin{enumerate}    
    \item Test whether $c =0$ mod $\frac{1}{2}$ by assuming $c =0$ and testing whether the assumption is consistent. $c = 0$ means the given $\Gth$-representation defined by $(\hS, \hT^2)$ is in fact a linear congruence represenatation of some level $N$.

    \begin{enumerate}
        \item First obtain the  level candidate $N$: from $(\hS, \hT^2)$ we can compute $T^{\rm ind}$ of Eq. \eqref{eq:ind_superMD}, 
        compute its eigenvalues $\lambda_i$, and find the smallest $N$ such that $\lambda_i^N = 1$ for all $i$. This $N$ will be the level candidate.

        \item With the  level candidate $N$, carry out the test of the relations Eq. \eqref{eq:congruence_hat} (note that the relations depend on $N$).
    \end{enumerate}
    If $(\hS, \hT^2)$ passes this test, then $c = 0$ mod $\frac{1}{2}$. Otherwise, $c \neq 0$ mod $\frac{1}{2}$ and there is no gapped boundary. 

    \item If we pass the test for $c = 0$ mod $\frac{1}{2}$,
    we compute $\hat{H}(n)$ for all $n$ such that ${\rm gcd}(n, N) = 1$ by Eq. \eqref{eq:hatHn0}. Then, the higher central charges for $(\breve{\cB}, 0)$ are given by  
     \begin{equation}
         \xi_n = \sum_{a} \hat{H}(n)_{1a} 
     \end{equation}
    which are guaranteed to be $\pm 1$ since $\hat{H}(n)$ are signed permutation matrices. We need all of them to be $+1$ for a gapped boundary to exist. 
\end{enumerate} 

\subsubsection*{Proof}

\emph{\textbf{ First step}}: Given a super-MTC $\cB$, the representation $\hat{\rho}$ of Eq. \eqref{eq:linear_rep_Gth} is a linear representation for any choice of $c$ consistent with the chiral central charge of $\cB$. In particular, if $\cB$ has $c = 0$ mod $\frac{1}{2}$, then $c = 0$ is a valid choice, and hence $\hat{\rho}$ given by
\begin{align}
    \hat{\rho}(s) := \hS \nonumber \\ 
    \hat{\rho}(t) := \hT
\end{align}
is a linear congruence representation of $\Gth$ of some level $N$. In order to determine $N$, we use the result of Sec. \ref{sec:order}, which states that $N$ is the order of the eigenvalues  $\lambda_i$ of $T^{\rm ind}$.

The modular extension $(\breve{\cB}, 0)$ has a linear representation $\rho$ of $\SL$ and $\ker \rho \geq \Gamma(N_{FS})$, where $N_{FS}$ is the order of the $T$-matrix of $(\breve{\cB}, 0)$.  
By the proofs of Theorem 3.1 of Ref. \cite{Bonderson_2018} and Theorem III.1 of Ref. \cite{cho2023classification},  we obtain
\begin{align}
    \ker \hat{\rho} \geq \Gamma(N_{FS}),
    \label{eq:kernel_larger}
\end{align}
i.e. the level of $\hat{\rho}$ is at most $N_{FS}$. On the other hand, by the results of Sec. \ref{sec:super_MTC}, $\rho_{\rm Ind}$ is a an $\SL$-representation of level $N_{FS}$, where $\rho_{\rm Ind}$ is a representation of $\SL$ induced from $\hat{\rho}.$   In fact, we think of $\hat{\rho}$ as a representation of $\Gth/\Gamma(m)$ for some even $m$, which then induces a representation $\rho_{\rm Ind}$ of $\SL/\Gamma(m).$ Then it is clear that $\ker \rho_{\rm Ind} \geq \Gamma(m)$. But $m$ cannot be any smaller than $N_{FS}$ (note that both $m$ and $N_{FS}$ are even).  Hence the level of $\hat{\rho}$ is at least $N_{FS}.$ Together, we see that the level of $\hat{\rho}$ is exactly $N_{FS}.$

$N_{FS}$ can be computed from the eigenvalues of $\rho_{\rm Ind}(t)$. Once we have the level candidate $N_{FS}$, we can simply check whether the relations Eq. \ref{eq:congruence_hat} are satisfied for $\hat{\rho}.$ 

\emph{\textbf{Second step}}: Since we only consider this step if the first step has been passed, we assume $c = 0 $ mod $\frac{1}{2}$ throughout this part.

 According to Ref. \cite{Kaidi_2022}, the higher central charges of an MTC are given by the chiral central charges of is Galois conjugates.  In particular, when $c = 0$ mod $8$,  $\xi_n$ can be computed as the phase of $S'(n)_{11} =  (H(\bar{n}) S)_{11}$  (see Eqs. (83) and (84) of Ref. \cite{Kaidi_2022}), where $S'(n)$ is the $S$-matrix of the Galois conjugate corresponding to $n \in \ZZ_{N_{FS}}^\times.$

We compute
    \begin{equation}
       S'(n)_{11} =  \sum_a   H(\bar{n})_{1a} S_{a1} = \sum_a   H(\bar{n})_{1a} \frac{d_a}{D}
       \label{eq:hcc_test}
    \end{equation}
   Note that $d_a$ and $D$ are real numbers (even for non-unitary MTCs), so $\xi_n = \frac{S'(n)_{11}}{|S'(n)_{11}|} = \pm 1$ (this is a general result that holds for higher central charges when $c=0$ mod $8$). 

    We can then compute the phase of Eq. \eqref{eq:hcc_test} for all $n \in \ZZ_{N_{FS}}^\times$. Any value of $-1$ will give an obstruction to the existence of a gapped boundary.  We now assume that the MTC we begin with is unitary, which means that $d_b$ are all positive. Then, 
    \begin{align}
        \xi_n =  \sum_b H(\bar{n})_{1b} = \pm  1
    \end{align}
where have used the fact that $H(\bar{n})$ is a signed permutation matrix. 

 Now, for a super-MTC $\cB$, the $\xi_n$ of the modular extension $(\breve{\cB}, 0)$ obey the above. We can restrict the $\SL$-representation formed by the modular data of $(\breve{\cB}, 0)$ to $\Gth$, after which $\rho$ becomes reducible with $\hat{\rho}$ as the first block. $H(n)$ survives the restriction, and is thus block-diagonalizable. We can write the first block purely in terms of the super-MTC data $\hS, \hT^2$ as (cf. Eq. \eqref{eq:hatHn})
 \begin{align}
       \hat{H}(n) :=& \hat{\rho}\begin{pmatrix}
        n & 0 \\ 0& \bar{n} 
    \end{pmatrix}   \nonumber \\ 
    =& \hS^2 \hT^{n^2-n} \hS \hT^{-(\bar{n}-1)} \hS (\hT^2 \hS)^{n-1}
    \label{eq:hatHn0}
 \end{align}
Note that because $N_{FS}$ is even and hence $n$ is odd, this expression only involves even powers of $\hT$ and is well-defined.  
$\hat{H}(n)$ is again a signed permutation matrix, as we demonstrate in Appendix \ref{app:signed_perm}. Moreover, again by Appendix \ref{app:signed_perm}, it is clear that the nonzero entry of the first row of  $\hat{H}(n)$ is equal to the nonzero entry of the first row of $H(n)$, i.e.
 \begin{equation}
     \sum_a \hat{H}_{1a} = \sum_a H_{1a}.
 \end{equation}
 Hence, we can compute the higher central charges of the modular extension $(\breve{\cB}, 0)$ purely in terms of the modular data of the super-MTC $\cB$ as 
 \begin{align}
     \xi_n =   \sum_a \hat{H}_{1a}.
 \end{align}\qedsymbol

\section{Examples}
\label{sec:examples}
We apply our test to known examples of super-MTCs to see which of them admit a gapped boundary. Ref. \cite{cho2023classification} classified super-MTCs up to rank 10. There are many super-MTCs with $c = 0$ mod $\frac{1}{2}$; however, we find that only the following (among unitary super-MTCs) pass the higher central charge test: 
\begin{itemize}
    \item $PSU(2)_{10}$ and $PSU(2)_{-10}$ (rank 6)
    \item $PSU(2)_6 \boxtimes_f PSU(2)_6$ (rank 8)
\end{itemize}
In particular, the new classes of rank $10$ modular data found by Ref. \cite{cho2023classification}, which were  constructed using the Drinfeld center of near-group fusion categories in Ref.  \cite{rowell2023neargroup}, do not admit gapped boundaries in spite of having $c = 0$ mod $\frac{1}{2}$.

Ref. \cite{Lan_2016theory} also lists several super-MTCs of rank 12 and 14.  Among these, we test those super-MTCs which are non-split with $c = 0$ mod $\frac{1}{2}$. 
\begin{itemize}
    \item For rank 12, one class of examples come from the fermion condensation of $U(1)_8 \boxtimes {\rm Ising}^{\nu}$ or similar \footnote{In this context, ``fermion conensation'' simply refers to obtaining the super-MTC from the spin-MTC/modular extension, rather than the more elaborate processes of Ref. \cite{Aasen_2019} or Ref. \cite{Delmastro_2021} by which one obtains a spin-TQFT}. These do not have vanishing $\xi_n$. 

    \item For rank 12, another class of examples come from fermion condensation of $(B_2)_2 \boxtimes U(1)_4$ or similar. These do not have vanishing $\xi_n.$

    \item For rank 14, we have a class of examples which come from the fermion condensation of ${\rm Ising}^{\nu_1} \boxtimes {\rm Ising}^{\nu_2} \boxtimes {\rm Ising}^{\nu_3}$. These have vanishing $\xi_n.$
    \end{itemize}

Moreover, $SU(2)_{4k+2}$ and $SO(4k+2)_2$ are known to be  infinite series of spin-MTCs, so their fermion condensations yield an super-MTCs \cite{zhang_thesis}. For $SU(2)_{4k+2}$, the $c \neq 0$ mod $\frac{1}{2}$ (except for $SU(2)_{10}$,  which was  considered earlier), so the existence of a gapped boundary is already obstructed by the chiral central charge. For $SO(4k+2)_2$,  we check up to $4k +2 = 68$ (we use Refs. \cite{gannon1998level, naidu2009finiteness} to compute the modular data). we find that only the following has vanishing higher central charges 
\begin{itemize}
    \item Fermion condensation of $SO(36)_2$ (rank 14).
\end{itemize}

In addition, for cases where the explicit modular extension data is available (e.g. via Ref. \cite{MEX}), we verify that  higher central charge computed via our method from $(\hat{S}, \hat{T}^2)$ agrees with the higher central charges computed from the $c = 0$ modular extension via the bosonic formula Eq. \eqref{eq:hcc}. 

\section{Discussion}
In this article, we have developed a method   which tests whether a given super-MTC $\cB$ 
 (1) has chiral central charge $c =0$ mod $\frac{1}{2}$ and (2) if so, whether the $c=0$ modular extension $(\breve{\cB}, 0)$ has vanishing higher central charges. The test only makes use of the modular data $(\hS, \hT^2)$ of $\cB$ as input. This gives a set of  obstructions to the existence of a gapped boundary for a super-MTC $\cB$ which can be efficiently computed, and we apply this to known examples of super-MTCs to rule out gapped boundaries for many of them. 

 In addition to providing necessary conditions for the existence of a gapped boundary, our results are significant because it represents a step toward computing the chiral central charge $c$ of a super-MTC $\cB$  in terms of its modular data. Thus far, the the computation of $c$ mod $\frac{1}{2}$ for a super-MTC relied on finding the modular extensions, which is a highly nontrivial process \cite{Lan_2016theory} (see also  Ref. \cite{Cano_2014} for a discussion of the abelian case).  Ref. \cite{Kobayashi_2022} has developed a Gauss-Milgram-like formula giving obstructions to the existence of a gapped boundary of  fermionic topoloigcal orders with $U(1)_f$-symmetry, but a formula for the general case is lacking.  The first part of our test, on the other hand, tests if a given $\cB$ has $c =0$ mod $\frac{1}{2}$ using congruence representation theory without any extra assumptions on $\cB$ and without needing to compute the modular extensions. While this only tests if $c =0$ mod $\frac{1}{2}$ and does not in general determine $c$ mod $\frac{1}{2}$ itself (it is not straightforward to use the congruence representation relations in general, because the  relations  depend on the level $N$ and $N$ in turns depends on $c$), we believe that further investigation of the matter from the direction of congruence representation could lead to an algorithm which computes $c$ (and possibly all higher central charges as well) in terms of the modular data of $\cB.$

\section*{Acknowledgements}
M.Y. would like to thank Hee-Cheol Kim and Donghae Seo for helpful discussions and comments. This paper is supported by Basic Science Research Institute Fund, whose NRF grant number is 2021R1A6A1A10042944.

\appendix

\section{Gapped boundaries of spin-TQFTs and  gapped boundaries of modular extensions}
\label{app:gapped_bc}

Recall that a spin-MTC $\breve{\cB}$ (which can be thought of as a modular extension of a super-MTC $\cB$) gives rise to a $2+1$d spin-TQFT  $\cT_f(\breve{\cB})$ \cite{Bhardwaj_2017, Delmastro_2021}. 
In this appendix, we show that, on the level of the torus, a gapped boundary for $\breve{\cB}$ gives rise to a gapped boundary for $\cT_f$ and vice versa.  

Consider a spin-MTC $\breve{\cB}$ which has a Lagrangian algebra $\cL = \bigoplus_a Z_a a$,  $a \in \breve{\cB}$ (for convenience, we will abuse notation and write $a \in \breve{\cB}$, which is understood to mean that $a$ belongs to the set of isomorphism classes of simple objects of $\breve{\cB}$).  The Lagrangian algebra has dimension $\dim \cL = \sum_{a \in \breve{\cB}} Z_a d_a = D$ where $D = \sqrt{\sum_{a \in \breve{\cB}} d_a^2}$ is the total quantum dimension of $\breve{\cB}.$ 
Following Ref. \cite{Kobayashi_2022}, we can think of the boundary condition on a torus as a state in the torus Hilbert space, by placing the $2+1$d TQFT on $T^2 \times [0,1]$ with the gapped boundary condition $\cL$ on $T^2 \times \{ 1\}$; then the state on the Hilbert space at $T^2 \times \{0 \}$ is given by $|\cL\rangle = \sum \limits_a Z_a |a \rangle $ (up to normalization) where $|a\rangle $ are basis states on the torus. 

Let us first derive some properties of $\cL$. Since $\cB$ is a spin-MTC, it has a distinguished fermion $f$. We can decompose the set of simple objects of $\breve{\cB}$ into $\breve{\cB}_{\rm NS} \oplus \breve{\cB}_{\rm R}$, where $a \in \breve{\cB}_{\rm NS}$ if $a$ has trivial mutual braiding with $f$, $S_{af} = + \frac{d_a}{D}$, and $a \in \breve{\cB}_{\rm R}$ if $a$ has nontrivial mutual braiding with $f$, $S_{af} = - \frac{d_a}{D}.$ Likewise, we can decompose $\cL$ into $\cL = \cL_{\rm NS}^1 \oplus \cL_{\rm R}^1$, where each condensing boson belongs to  $\cL_{\rm NS}$ ($\cL_{\rm R}$) if it braids trivially (nontrivially) with $f.$

 Since the boundary is  topological, it is invariant under modular transformations; this leads to $Z_a$ being invariant under $S$- and $T$-transformations \cite{Ji_2019, Kaidi_2022}. Invariance under $T$ simply means the anyons $a$ for which $Z_a$ are nonzero are all bosons. Invariance under $S$, $\sum_b S_{ab} Z_b = Z_a$, puts additional constraints. Let us specialize to $a = f$, the simple object representing the fundamental fermion. Then we see 
\begin{equation}
     \sum_{b \in \breve{\cB}} S_{fb} Z_b 
 = \frac{1}{D} \left( \sum_{b \in \breve{\cB}_{\rm NS}}  d_b Z_b - \sum_{b \in \breve{\cB}_{\rm R}}  d_b Z_b \right) = Z_f = 0   
\end{equation}
since $f$, as a fermion, does not belong to the Lagrangian algebra (enters with coefficient $Z_f = 0$). This leads to 
\begin{equation}
     \sum_{b \in \breve{\cB}_{\rm NS}}  d_b Z_b  =  \sum_{b \in \breve{\cB}_{\rm R}}  d_b Z_b
\end{equation}
which means that $\cL_{\rm NS}^0$  and $\cL_{\rm R}^0$ have the same quantum dimensions; each has $\frac{1}{2} D$. 

We can then define
\begin{align}
    \cL_{\rm NS}^f = f \times \cL_{\rm NS}^1  \\ 
    \cL_{\rm R}^f= f \times \cL_{\rm R}^1
\end{align}
and 
\begin{align}
     \cL_{\rm NS} =  \cL_{\rm NS}^1  \oplus \cL_{\rm NS}^f\\ 
   \cL_{\rm R} =  \cL_{\rm R}^1  \oplus \cL_{\rm R}^f.
\end{align}
This gives us the NS and R sector Lagrangian algebras of Ref. \cite{Kobayashi_2022}, but generalized to the case where  $\breve{\cB}_{\rm R}$ possibly contains ``q-type'' anyons which absorb  the fermion, $q \times f = q.$ Note that $\cL_{\rm NS}^f$ is always distinct from $\cL_{\rm NS}^1$, as none of the anyons in $\cB_{\rm NS}$ absorb $f$. On the other hand, if $\cL_{\rm R}^1$ contains a $q$-type anyon, it will also appear in $\cL_{\rm R}^f$; in such cases, it will appear in $\cL_{\rm R}$ with multiplicity.

To consider the spin-TQFT, we change the basis into the ``sector basis,'' as in Eq. \eqref{eq:basis}.  This amounts to considering the states 
\begin{align}
    |\cL_{\text{NS-NS}} \rangle = \sum_a (Z_{\rm NS,a}^1 + Z_{\rm NS,a}^f) |a \rangle  \\
    |\cL_{\text{NS-R}} \rangle = \sum_a (Z_{\rm NS,a}^1 - Z_{\rm NS,a}^f) |a \rangle \\
    |\cL_{\text{R-NS}} \rangle = \sum_a (Z_{\rm R,a}^1 + Z_{\rm R,a}^f) |a \rangle \\
    |\cL_{\text{R-R}} \rangle = \sum_a (Z_{\rm R,a}^1 - Z_{\rm R,a}^f) |a \rangle. 
\end{align}

The R-R sector of the $2+1d$ spin-TQFT also has puncture states $|q; f \rangle$ for each $q$ which absorbs a fermion \cite{Delmastro_2021}. However, to describe a gapped boundary of a $2+1$d spin-TQFT on the torus, i.e. a $1+1$d spin-TQFT, it is sufficient to give its partition functions (more precisely, these are partition functions on tori with specified spin structures and with  ``background field'' specified by the anyon $a$ \cite{Ji_2019}):
\begin{align}
    Z_{\text{NS-NS}, a} = Z_{\rm NS,a}^1 + Z_{\rm NS,a}^f  \\
    Z_{\text{NS-R}, a}  = Z_{\rm NS,a}^1 - Z_{\rm NS,a}^f\\
    Z_{\text{R-NS}, a}  = Z_{\rm R,a}^1 + Z_{\rm R,a}^f \\
    Z_{\text{R-R}, a} = Z_{\rm R,a}^1 - Z_{\rm R,a}^f. 
\end{align}

How do these transform under modular transformations? Using the invariance of $Z_a = Z_{{\rm NS}, a}^1 + Z_{{\rm R}, a}^1$ under $S$, together with the following facts: $Z_{{\rm NS},a}^1 = 0$ for $a \in \breve{\cB}_{\rm R}$ and $Z_{{\rm R},a}^1 = 0$ for $a \in \breve{\cB}_{\rm NS}$;  $S_{a, b \times f} = \epsilon_a S_{ab}$ where $\epsilon_ a = +1$ if $a \in \breve{\cB}_{\rm NS}$ and $-1$ if $a \in \breve{\cB}_{\rm R}$; $Z_{{\rm NS}, a}^f = Z_{{\rm NS}, a \times f}^1$ and $Z_{{\rm R}, a}^f = Z_{{\rm R}, a \times f}^1$, we compute
\begin{align}
 \sum_{b \in \breve{\cB}} S_{ab} \left(   Z_{{\rm NS},b}^1 + Z_{{\rm R},b}^1 \right) &= Z_{ {\rm NS},a}^1 + Z_{{\rm R},a}^1 \\
  \sum_{b \in \breve{\cB}} S_{a\times f , b} \left(   Z_{{\rm NS},b}^1 + Z_{{\rm R},b}^1 \right) &=\sum_{b \in \breve{\cB}} S_{a b} \left(   Z_{{\rm NS},b}^1 - Z_{{\rm R},b}^1 \right) \nonumber \\ 
  = Z_{ {\rm NS},a}^f + Z_{{\rm R},a}^f \\
    \sum_{b \in \breve{\cB}} S_{a , b \times f} \left(   Z_{{\rm NS},b}^1 + Z_{{\rm R},b}^1 \right) &= 
    \sum_{b \in \breve{\cB}} S_{ab} \left(   Z_{{\rm NS},b}^f + Z_{{\rm R},b}^f \right) \nonumber \\
    = Z_{ {\rm NS},a}^1 - Z_{{\rm R},a}^1 \\
    \sum_{b \in \breve{\cB}} S_{a\times f , b \times f} \left(   Z_{{\rm NS},b}^1 + Z_{{\rm R},b}^1 \right) &=\sum_{b \in \breve{\cB}} S_{ab} \left(   Z_{{\rm NS},b}^f - Z_{{\rm R},b}^f \right)   \nonumber \\ 
    =Z_{ {\rm NS},a}^f - Z_{{\rm R},a}^f
\end{align}

From this we can see that 
\begin{align}
    \sum_{b \in \breve{\cB}} S_{ab}  Z_{\text{NS-NS}, b} =  Z_{\text{NS-NS}, a}
    \label{eq:S_NSNS}\\
    \sum_{b \in \breve{\cB}} S_{ab}  Z_{\text{NS-R}, b} =  Z_{\text{R-NS}, a} \label{eq:S_NSR}\\
    \sum_{b \in \breve{\cB}} S_{ab}  Z_{\text{R-NS}, b} =  Z_{\text{NS-R}, a} \label{eq:S_RNS}\\
    \sum_{b \in \breve{\cB}} S_{ab}  Z_{\text{R-R}, b} =  Z_{\text{R-R}, a}.
    \label{eq:S_RR}
\end{align}
Moreover, from the fact that   $Z_{\text{NS}, a}^f$  is nonzero only for $a$ such that $\theta_a = -1$, while $Z_{\text{NS}, a}^1$, $Z_{\text{R}, a}^1$ and $Z_{\text{R}, a}^f$ nonzero only for $a$ such that $\theta_a = 1$,  it can easily be seen that
\begin{align}
    \sum_{b \in \breve{\cB}} T_{ab}  Z_{\text{NS-NS}, b} =  Z_{\text{NS-R}, a} \\
    \sum_{b \in \breve{\cB}} T_{ab}  Z_{\text{NS-R}, b} =  Z_{\text{NS-NS}, a} \\
    \sum_{b \in \breve{\cB}} T_{ab}  Z_{\text{R-NS}, b} =  Z_{\text{R-NS}, a} \\
    \sum_{b \in \breve{\cB}} T_{ab}  Z_{\text{R-R}, b} =  Z_{\text{R-R}, a}.
    \label{eq:T_RR}
\end{align}
Combined, these equations show that $S$ maps between the NS-R sector to the R-NS sector while keeping the NS-NS and R-R sectors invariant, while $T$ maps between the NS-NS and NS-R sectors while keeping the R-NS and R-R sectors invariant. This is exactly the behavior of the partition functions of a $1+1$d spin-TQFT. Thus, we have shown that a gapped boundary  $\cL$ of $\breve{\cB}$ gives rise to a gapped boundary on the torus for the corresponding $2+1$d spin-TQFT $\cT_f(\breve{\cB})$. We can also re-phrase this as: the partition function of each sector is invariant under a subgroup of $\SL$ \cite{Delmastro_2021}:
\begin{itemize}
    \item NS-NS: $\Gth = \langle s, t^2 \rangle $
    \item NS-R: $\Gamma^0(2) = \langle tst, t^2 \rangle $
    \item R-NS: $\Gamma_0(2) = \langle st^2s, t \rangle $
    \item R-R: $\SL$ itself.
\end{itemize}

On the other hand, suppose we are given a gapped boundary for a $2+1$d spin-TQFT $\cT_f(\breve{\cB})$ on the torus, i.e. a set of partition functions $ Z_{\text{NS-NS}, a}$, $ Z_{\text{NS-R}, a}$,  $Z_{\text{R-NS}, a}$, and $ Z_{\text{R-R}, a}$ which satisfy Eqs. \eqref{eq:S_NSNS} through \eqref{eq:T_RR}. Then we can construct the gapped boundary for the bosonic TQFT corresponding to $\breve{\cB}$ as follows:
\begin{equation}
    Z_a =  \frac{1}{2} \left( Z_{\text{NS-NS}, a} +  Z_{\text{NS-R}, a} +  Z_{\text{R-NS}, a} +  Z_{\text{R-R}, a} \right)
\end{equation}
  This is indeed invariant under $S$ and $T$ of $\breve{\cB}.$ 
Note that invariance under $S$, together with $Z_1 = 1$ automatically means that we have the correct total quantum dimension, since
\begin{equation}
   D \sum_a S_{1a} Z_a =\sum_a d_a Z_a =  \dim \cL =  D Z_1 = D. 
\end{equation}

\section{Higher central charges for super-MTC}
\label{app:hcc_defined}
Going beyond the $c= 0$ case, we show that higher central charges $\xi_n$ for odd $n$ are well-defined up to factors of $e^{2 \pi i /16}$ for super-MTCs (similar to how  the chiral central charge $c$ is well-defined modulo $\frac{1}{2}$). Note that the presence of a fermion with $\theta_f = -1$ means that $N_{FS}$ is always even for a super-MTC; hence any $n$ coprime to $N_{FS}$ is odd, and restricting to odd $n$ loses no value as far as the relevance to gapped boundaries is concerned.  

First, we note that if $\breve{\cB}$ is a modular extension of a super-MTC $\cB$, other modular extensions of $\cB$ can be constructed by stacking $\breve{\cB}$ with $\cS_{\nu}$, one of the Kitaev 16-fold way phases \cite{Kitaev_2006} ,  and then condensing the boson formed by $(f, \psi)$ where $f$ is the distinguished fermion of $\breve{\cB}$ and $\psi$ is the distinguished fermion of $\cS_\nu.$  
We denote the resulting  theory as $\left( \breve{\cB} \boxtimes \cS_{\nu} \right)_{\ZZ_2}.$

Higher central charges simply multiply under stacking. If we are given two MTCs $\cB$ and $\cD$,
\begin{align}
     (\xi_n)_{\cB \boxtimes \cD}=   \frac{\sum \limits_{b \in \cD} \sum \limits_{a \in \cB} (d_a d_b)^2 (\theta_a \theta_b)^{n}}{|\sum \limits_{b \in \cD} \sum \limits_{a \in \cB} (d_a d_b)^2 (\theta_a \theta_b)^{n}|} \nonumber \\
     = \left(\frac{ \sum_a d_a^2 \theta_a^n }{| \sum_a d_a^2 \theta_a^n |}\right) \left( \frac{\sum_b d_a^2 \theta_b^n}{|\sum_b d_a^2 \theta_b^n|} \right)   = (\xi_n)_{\cB}(\xi_n)_{\cD}.
\end{align}
This means that, under stacking with $\cS$, the higher central charges $(\xi_n)_{\breve{\cB}}$ of $\breve{\cB}$  change by the higher central charges $(\xi_n)_{\cS_\nu}$ of $\cS_\nu$. $\cS_\nu$ are well-known, and we can easily compute their higher central charges. We do not list them explicitly, but simply note that they are all 16th roots of unity. 

Now, we show that under condensation of a $\ZZ_2$ boson, $\xi_n$ remains invariant for odd $n.$ Suppose an MTC $\cC$ has  a $\ZZ_2$-boson $e$ (i.e. a boson such that $e \times e = 1$ and $\theta_e = 1)$). Then, every anyon $a \in \cC$ has mutual braiding phase $\pm 1$ with $e$, and we can divide $\cC$ in to two sectors $\cC_0 \oplus \cC_1$ where anyons in $\cC_0$  have trivial braiding with $e$ and anyons in $\cC_1$ have $-1$ braiding with $e$. Because of the $\ZZ_2$-fusion rule of $e$, a given anyon $a$ may either absorb $e$ or not under fusion: $a \times e = a$ (``short orbit'') or $a \times e = a^e \neq a$ (``long orbit''). If $a$ absorbs $e$, it has to belong to the $\cC_0$ sector (see Appendix A.1 of Ref. \cite{Delmastro_2021}). Hence every anyon $a \in \cC_1$ comes in pairs $a, a^e$, and moreover they satisfy $\theta_{a^e} = - \theta_a.$ This means that
\begin{align}
    \sum_{a \in \cC} d_a^2 \theta_a^n = \sum_{a \in \cC_0} d_a^2 \theta_a^n  + \sum_{a \in \cC_1^1} d_a^2 \theta_a^n  + \sum_{a \in \cC_1^e} d_a^2 \theta_a^n  \nonumber \\ 
    = \sum_{a \in \cC_0} d_a^2 \theta_a^n  + \sum_{a \in \cC_1^1} d_a^2 \theta_a^n  - \sum_{a \in \cC_1^1} d_a^2 \theta_a^n = \sum_{a \in \cC_0} d_a^2 \theta_a^n  
   \label{eq:hcc_Z2_condense}
\end{align}
for $n$ odd. 

Condensation corresponds to throwing away the anyons in $\cC_1$ (which are confined), and then identifying anyons related by fusion with $e$, and splitting anyons which absorb $e$. We already see from Eq. \eqref{eq:hcc_Z2_condense} that confined anyons do not affect $\xi_n$ for odd $n$. Identifying $a \sim a \times e$ means that long orbits now contribute half as much to the sum; splitting the short orbit means that instead of $d_a^2 \theta_a$, it now contributes $2 (\frac{1}{2}d_a)^2 \theta_a$ so again  half as much. So, after condensation, we simply get an overall factor of $\frac{1}{2}$ in front of the sum $\sum_{a \in \cC}d_a^2 \theta_a$. Since $\xi_n$ is the phase of this, this factor does not affect it.

Thus, $\left( \breve{\cB} \boxtimes \cS_{\nu} \right)_{\ZZ_2}$ has higher central charges which differ from those of $\breve{\cB}$ by at most factors of 16th roots of unity, and higher central charges of $\cB$ are well-defined up to factors of 16th roots of unity.

\section{Proof of that \texorpdfstring{$\rho_I$}{} is an induced representation of \texorpdfstring{$\hat{\rho}$}{}  }
\label{app:induced}
Here we prove a result used in Sec. \ref{sec:order}.

Recall that $\rho_I$ takes the following form (cf. Eqs \eqref{eq:spinsector}):
\begin{align}
        \rho_I(s) &= 
    \begin{pmatrix}
        \hat{\rho}(s) & 0 & 0 \\
        0 & 0& \Tilde{S}' \\
        0 & \Tilde{S}'^T & 0    
    \end{pmatrix}, \nonumber \\
    \rho_I(t) &= \begin{pmatrix} 0 & \hat{\rho}(t)  & 0  \\ \hat{\rho}(t) & 0&  0  \\ 0 & 0 & \Tilde{T}'.  \end{pmatrix}
    \label{eq:spinsector_lin}
\end{align}
If we restrict this to $\Gth$, the first block becomes decomposable and gives us exactly $\hat{\rho}$. Here,  $\Tilde{S}'$ and $\Tilde{T}'$ are some matrices whose details do not matter. The crucial fact is that $\rho_I$ takes this block form, where it 
 acts on three subspaces $V_1, V_2, V_3$ by mapping among them. If $\rho_I$ is  a direct sum $\bigoplus_i (\rho_I)_i$ of irreps $(\rho_I)_i$, each of the summands $(\rho_I)_i$ also has to map among the three subspaces, and hence are of dimension $3 k_i$, $k_i \in \mathbb{N}$. If we restrict to $\Gth$, $\rho_I |_{\Gth}$ and hence each $(\rho_I)_i|_{\Gth}$ leaves the first subspace $V_1$ invariant. Each of $(\rho_I)_i|_{\Gth}$ then contributes  a $k_i$-dimensional irrep $\hat{\rho}_i$ to the $\Gth$-representation $\hat{\rho}$ which acts on $V_1$ ($\hat{\rho}_i$ has to be an irrep because, by a straightforward application of Frobenius reciprocity, the restriction of a  $G$-representation $R$ to $H$ can only contain $H$-irreps of dimension greater than or equal to  $\frac{1}{[G:H]} \dim R$). Hence $\hat{\rho} = \bigoplus_i \hat{\rho}_i$ \footnote{We refer to Ref. \cite{cho2023classification} for an exposition of restriction and induction of representations, as well as Frobenius reciprocity, in the context of $\SL$ and its subgroup $\Gth$.}.

Now, since each $\hat{\rho}_i$ and $(\rho_I)_i$ are irreps, we get ${\rm Ind} \hat{\rho}_i = (\rho_I)_i$ by Frobenius reciprocity.  Then, 
\begin{align}
    {\rm Ind} \hat{\rho} = \bigoplus_i {\rm Ind} \hat{\rho}_i = \bigoplus_i  (\rho_I)_i = \rho_I
\end{align}
which shows that $\rho_I$ is indeed the induced representation of $\hat{\rho}$. \qedsymbol

\section{Proof that \texorpdfstring{$\hat{H}(n)$}{} is a signed permutation matrix}
\label{app:signed_perm}

From Sec. \ref{sec:congruence} and  \ref{sec:super_MTC}, we know that $H(n) := \rho\begin{pmatrix} n & 0 \\ 0& \bar{n} \end{pmatrix}$, $n \in \ZZ_n^{\times}$, is a  symmetric signed permutation matrix. 


After a basis change,  $U\rho U^{\dagger} = \hat{\rho} \oplus ...$, where $U$ is explicitly
\begin{align}
    U = \frac{1}{\sqrt{2}}\begin{pmatrix} \mathds{1} & \mathds{1} \\ \mathds{1} & - \mathds{1}\end{pmatrix} \oplus\frac{1}{\sqrt{2}}\begin{pmatrix} \mathds{1} & \mathds{1} \\ \mathds{1} & - \mathds{1}\end{pmatrix} \oplus \mathds{1}.
\end{align}
(This $U$ maps between Eqs. \eqref{eq:MEX_ST} and \eqref{eq:spinsector}.)

From this it follows that $UH(n)U^{\dagger} = \hat{H}(n) \oplus ...$ since $\hat{H}(n) := \hat{\rho}\begin{pmatrix} n & 0 \\ 0& \bar{n} \end{pmatrix}$. 

Let us write 
\begin{align}
    H(n) = \begin{pmatrix} A & B & \cdots  \\ C & D & \cdots \\ \vdots  & \vdots & \ddots \end{pmatrix}
\end{align}
where $A, B, C, D$ are each $d$-dimensional matrices, and $B = C^T, \ A^T = A, \ D^T = D$ because $H(n)$ is symmetric. Then we have
\begin{align}
     \begin{pmatrix} \hat{H}(n) & 0 & \cdots  \\ 0  & \ddots & \cdots \\ \vdots  & \vdots & \ddots  \end{pmatrix} = UH(n)U^{\dagger} \nonumber \\ 
    = \frac{1}{2}\begin{pmatrix} A+B+C+D & A+C-(B+D) & \cdots  \\ A-C + B-D & A-C -(B-D) & \cdots \\ \vdots  & \vdots & \ddots  \end{pmatrix}
\end{align}
from which we get $\hat{H}(n) = \frac{1}{2} (A+B+C+D)$ and $0 = A+C - B - D = A- C +B - D$, from which we see that \begin{align}
    B = C
\end{align}
 and \begin{align}
     A - D = 0.
 \end{align}
Then, 
\begin{equation}
    \hat{H}(n)= A + B.
\end{equation}
 Now, we note that $A, B$ are part of $H(n)$, a signed permutation matrix, on the same block row. If an entry $A_{ij}  = \pm 1$, then the every entry of $H(n)$ on the same row should be $0$. Hence,  $B_{ik} = 0$ for all $k$. It is also clear that if  $A_{ij} = \pm 1$, $A_{ik} = 0$ for all $k \neq j$. Then, $A+B$ has no row with multiple nonzero entries. 
 
 Similarly, since $C = B$, we can repeat the same analysis for columns, and show that $A+B$ has no column with multiple nonzero entries. Moreover, since $A_{ij}$ and $B_{ij}$ cannot both be nonzero for any $i, j,$ every entry of $A+B$ is $1,\  0,$ or $-1$. 
 
We know $\emph{a priori}$ that $\hat{H}(n)$ must be unitary. 
Hence, $\hat{H}(n) = A +B$ is a unitary matrix whose entries are $1, \ 0,$ or $-1$ and with at most one  nonzero entry per each row or column -- i.e. it is a signed permutation matrix.  \qedsymbol

Moreover, this means that the first row of $A +B$ contains a nonzero entry $\pm 1$, and this value must equal the value of the nonzero entry of the first row of $H(n)$.

\bibliographystyle{apsrev4-2}
\bibliography{bib.bib}

\end{document}